\documentclass{IET}

\begin{document}

\title{Degrees of Freedom of $M\times N$ SISO X Channel with Synergistic Alternating CSIT}

\author[1]{Young-Sik Moon}
\affil{Department of Electrical and Computer Engineering, INMC, Seoul National University, Seoul 151-744, Korea}

\author[2]{Jae-Hong Kim}

\affil{Samsung Electronics, Company, Ltd., Gyeonggi-do, Korea}

\author[1]{Jong-Seon No}
\author[3]{Dong-Joon Shin}
\affil{Department of Electronic Engineering, Hanyang University, Seoul 133-791, Korea}

\affil[*]{myskill@ccl.snu.ac.kr}

\abstract{In this paper, degrees of freedom (DoF) is investigated for the $M\times N$ single input single output (SISO) X channel with alternating channel state information at the transmitters (CSIT). It is known that the sum DoF of 2-user SISO X channel with synergistic alternating CSIT is the same as the sum DoF of 2-user $(M=N=2)$ SISO X channel with perfect CSIT [8]. In this paper, such 2-user X channel schemes are extended to the general $M\times N$ X channel. It is shown that the proposed $M\times N$ X channel schemes with synergistic alternating CSIT achieve $2M/(M+1)$ sum DoF. This DoF with $M=N=K$ is strictly lager than the best known DoF for the $K$-user X channel with delayed CSIT.}

\maketitle

\section{Introduction}
Interference alignment is an important technique to manage interference in the wireless communication networks. Most of the previous works on the interference alignment [1], [2] assume that perfect channel state information (CSI) is available at all transmitters. However, in practical communication environment, it is very difficult for transmitters to use the perfect CSI because CSI should be instantaneously available at the transmitter with no error. Thus, it is desirable to minimize the required channel state information at the transmitter (CSIT). Maddah-Ali and Tse [3] proved that completely outdated CSIT can still be useful even if the channel states are completely independent. In the multi-user wireless communication systems, some of channels are quasi-static, that is, CSI of those channels does not change over several time slots. Then the delayed CSIT can be considered as an instantaneous perfect CSIT in the next time slots. Thus, both the delayed and the instantaneous perfect CSITs can be assumed to coexist in the real multi-user wireless communication systems. Tandon et al. [7] suggested an alternating CSIT model in the broadcasting channel, where the availability of the delayed and the perfect CSITs varies over time. Recently, Wagdyy et al. [8], [9] introduced new achievable schemes for 2-user single input single output (SISO) X channel with alternating CSIT. These schemes achieve $4/3$ sum degrees of freedom (DoF), which is the theoretical upper bound for 2-user SISO X channel with instantaneous perfect CSIT. In this paper, based on the results in [8], [9], an achievable scheme is proposed for the $M\times N$ SISO X channel with synergistic alternating CSIT. Also, sum DoF for the ${M\times N}$ SISO X channel with synergistic alternating CSIT is derived.\\\indent The rest of this paper is organized as follows: In Section II, the system model is described. Then, an achievable scheme for $M\times N$ SISO X channel is proposed and its sum DoF is drived in Section III. Finally, conclusion is given in Section IV.

\section{System Model}
 We consider $M\times N$ SISO X channel, where each transmitter $T_j$, $j=1,...,M,$ transmits independent message $W_{ij}$ to each receiver $R_i$, $i=1,...,N$. Let $X_j(t)=f_{1j}(t)W_{1j}(t)+f_{2j}(t)W_{2j}(t)+\cdot\cdot\cdot+f_{Nj}(t)W_{Nj}(t)$ be the signal transmitted from the $j$th transmitter $T_j$ at the time slot $t$, where $f_{ij}(t)$ is the precoding coefficient for the message $W_{ij}(t)$. The received signal at the $i$th receiver at the time slot $t$ is given as
\begin{align}
Y_{i}(t)=\sum_{j=1}^M h_{ij}(t)X_{j}(t)+N_{i}(t)     
\end{align}
where $h_{ij}(t)$ denotes the channel coefficient between the $j$th transmitter and the $i$th receiver and $N_i(t)\sim\mathcal{CN}(0,1)$ is the circularly symmetric white Gaussian noise with zero mean and unit variance at the receiver $R_i$.
The power constraint is assumed to be $E||X(t)||^2 \leq P$. 
Let $r_{ij}(P)=\log_2|W_{ij}|/n$ denote the achievable rate per channel use of $W_{ij}$ for the transmission power $P$, where $|W_{ij}|$ denotes the alphabet size of $W_{ij}$. The DoF region $\mathcal D$ of the $M\times N$ X channel is defined as the set of all real non-negative tuples $(d_{11},d_{12},...,d_{MN})\in R_{+}^{MN}$, where $d_{ij}=\lim_{P\rightarrow\infty}  {{r_{ij}(P)}/{\log P}}$. The sum DoF of $M\times N$ X channel is defined as [1]
\begin{align}
\text {DoF}=\max_{(d_{11},d_{12},...,d_{MN})\in \mathcal D}\sum_{i=1}^{M}\sum_{j=1}^{N}d_{ij}.  
\end{align}
\indent It is considered that there are three different states of availability of CSIT for the receiver side as [8]:\\
1) Perfect CSIT (P state): CSIT is available to the transmitters instantaneously and without error.\\
2) Delayed CSIT (D state): CSIT is available to the transmitters with some delay 
and without error.\\
3) No CSIT (N state): CSIT is not available to the transmitters at all.\\
If the receiver $R_i$ is in the P state, then each transmitter knows the perfect CSI $h_{ij}(t)$, $1\leq j \leq M$ at time slot $t$. In the same sense, if the receiver $R_i$ is in the D state, then each transmitter knows the delayed CSI $h_{ij}(t)$, $1\leq j \leq M$ at time slot $t+\tau$ with the delay $\tau$.  

\section{Proposed Achievable Scheme}
In this section, we propose achievable schemes for the 3-user SISO X channel $(M=N=3)$ and extend them to the general $M\times N$ SISO X channel.
\subsection{Achievable Schemes for the 3-user SISO X Channel}
\begin{figure}[h]
\centering
\includegraphics[width=0.6\textwidth]{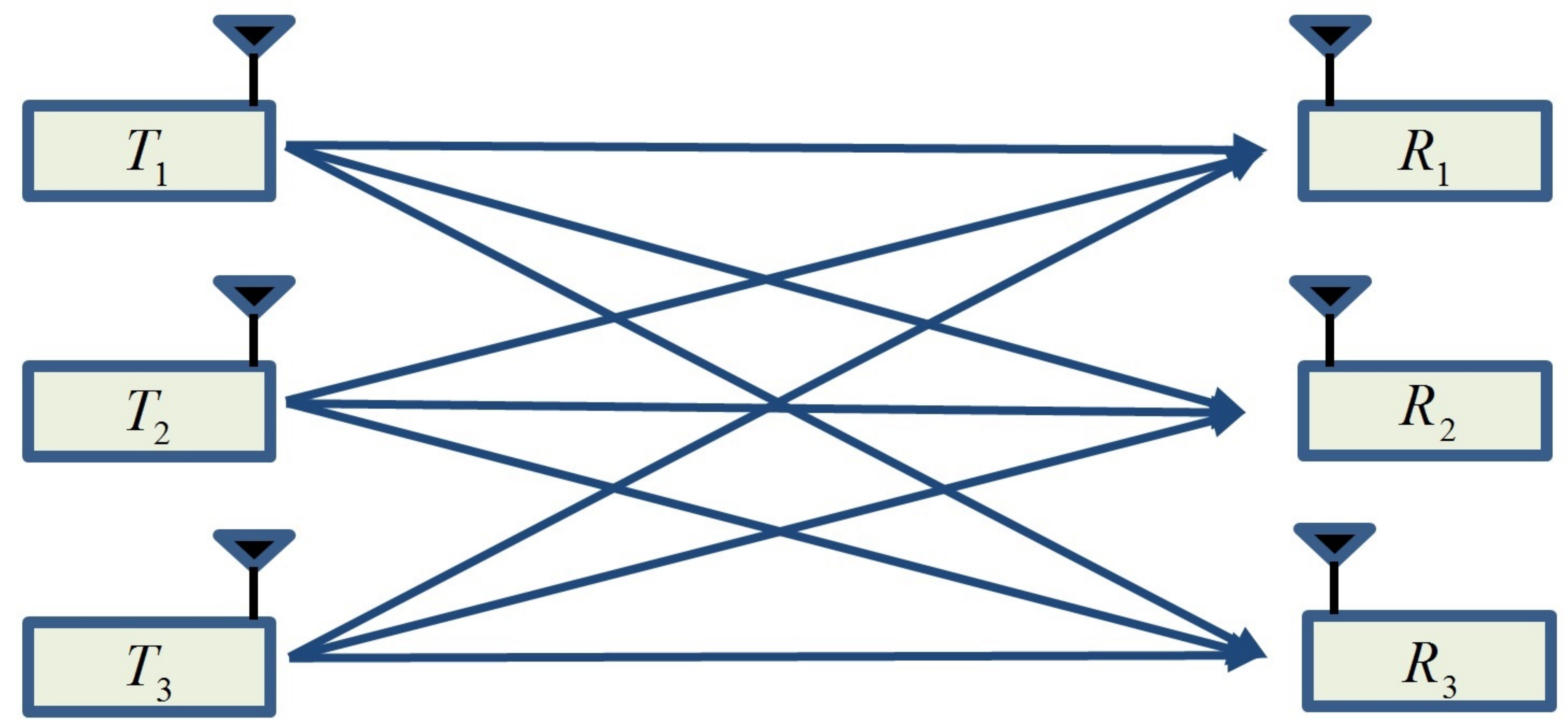}
\caption{3-user SISO X channel.}
\label{fig:}
\end{figure}
First, we extend the achievable schemes in [8] to the 3-user SISO X channel in Fig. 1. Using synergistic alternating CSIT in [8], $4/3$ DoF for the 2-user SISO X channel is achieved. 
In the proposed scheme for the 3-user case, each transmitter transmits independent messages to three receivers over six time slots. Similar to [8], the proposed achievable scheme for the 3-user case is composed of two separate phases described as below.\\\
 
\textit{\textbf{Phase 1)}}~~In this phase, each transmitter transmits its messages during three time slots. At time slot $t=i$, $1\leq i \leq 3$, all three transmitters transmit their messages for the receiver $R_i$, i.e., $X_j(t=i)=W_{ij},~1\leq i,j \leq 3$. Then, the transmitted signals at time slot $t$, $t=1, 2, 3$, are given as
\begin{align}
&X_1(1)=W_{11},~ X_2(1)=W_{12},~ X_3(1)=W_{13}\nonumber\\
&X_1(2)=W_{21},~ X_2(2)=W_{22},~ X_3(2)=W_{23}\\
&X_1(3)=W_{31},~ X_2(3)=W_{32},~ X_3(3)=W_{33}.\nonumber
\end{align}
Also, the received signals $Y_{i}(t),~1\leq i,t \leq 3$, at the receiver $R_i$ are given as
\begin{align}
&Y_{1}(1)=h_{11}(1)W_{11}+h_{12}(1)W_{12}+h_{13}(1)W_{13}\equiv L_1^1(W_{11},W_{12},W_{13})\nonumber\\
&Y_{2}(1)=h_{21}(1)W_{11}+h_{22}(1)W_{12}+h_{23}(1)W_{13}\equiv I_2^1(W_{11},W_{12},W_{13})\\
&Y_{3}(1)=h_{31}(1)W_{11}+h_{32}(1)W_{12}+h_{33}(1)W_{13}\equiv I_3^1(W_{11},W_{12},W_{13})\nonumber\\\nonumber\\\nonumber
&Y_{1}(2)=h_{11}(2)W_{21}+h_{12}(2)W_{22}+h_{13}(2)W_{23}\equiv I_1^1(W_{21},W_{22},W_{23})\nonumber\\
&Y_{2}(2)=h_{21}(2)W_{21}+h_{22}(2)W_{22}+h_{23}(2)W_{23}\equiv L_2^1(W_{21},W_{22},W_{23})\\
&Y_{3}(2)=h_{31}(2)W_{21}+h_{32}(2)W_{22}+h_{33}(2)W_{23}\equiv I_3^2(W_{21},W_{22},W_{23})\nonumber\\\nonumber\\\nonumber
&Y_{1}(3)=h_{11}(3)W_{31}+h_{12}(3)W_{32}+h_{13}(3)W_{33}\equiv I_1^2(W_{31},W_{32},W_{33})\nonumber\\
&Y_{2}(3)=h_{21}(3)W_{31}+h_{22}(3)W_{32}+h_{23}(3)W_{33}\equiv I_2^2(W_{31},W_{32},W_{33})\\
&Y_{3}(3)=h_{31}(3)W_{31}+h_{32}(3)W_{32}+h_{33}(3)W_{33}\equiv L_3^1(W_{31},W_{32},W_{33}),\nonumber
\end{align}
where $L_i^k$ denotes the $k$th linear combination of three desired messages $(W_{i1},W_{i2},W_{i3})$ for the receiver $R_i$ and $I_i^k$ denotes the $k$th interference signal to the receiver $R_i$, which is composed of three interference messages.\\\
 
\textit{\textbf{Phase 2)}}~~In this phase, all three transmitters transmit their messages for a pair of receivers at each time slot. Since there are total three receiver pairs, three time slots are needed in this phase. Note that the transmitted signal at the transmitter $j$ is precoded for interference cancellation by using the messages $\{W_{1j}, W_{2j}, W_{3j}\}$ and the delayed CSIs in the phase 1 and the perfect CSIs. In detail, at time slot $t=4$, $R_1$ and $R_2$ are P state, i.e., $T_j$ knows $h_{1j}(4)$ and $h_{2j}(4)$. Also, $R_1$ is D state at time slot $t=2$ and $R_2$ is D state at time slot $t=1$, i.e., $T_j$ knows $h_{1j}(2)$ and $h_{2j}(1)$. By using these CSIs, each transmitter transmits the precoded signal at time slot $t=4$ as follows.
\begin{align}
&X_{1}(4)=h_{21}^{-1}(4)h_{21}(1)W_{11}+h_{11}^{-1}(4)h_{11}(2)W_{21}\nonumber\\
&X_{2}(4)=h_{22}^{-1}(4)h_{22}(1)W_{12}+h_{12}^{-1}(4)h_{12}(2)W_{22}\\
&X_{3}(4)=h_{23}^{-1}(4)h_{23}(1)W_{13}+h_{13}^{-1}(4)h_{13}(2)W_{23}.\nonumber
\end{align}
Then, the received signals at time slot $t=4$ are given as
\begin{align}
Y_{1}(4)=&h_{11}(4)h_{21}^{-1}(4)h_{21}(1)W_{11}+h_{12}(4)h_{22}^{-1}(4)h_{22}(1)W_{12}+h_{13}(4)h_{23}^{-1}(4)h_{23}(1)W_{13}\nonumber\\&+h_{11}(2)W_{21}+h_{12}(2)W_{22}+h_{13}(2)W_{23}\nonumber\\\equiv& L_1^2(W_{11},W_{12},W_{13})+I_1^1(W_{21},W_{22},W_{23})\nonumber\\
Y_{2}(4)=&h_{21}(4)h_{11}^{-1}(4)h_{11}(2)W_{21}+h_{22}(4)h_{12}^{-1}(4)h_{12}(2)W_{22}+h_{23}(4)h_{13}^{-1}(4)h_{13}(2)W_{23}\\&+h_{21}(1)W_{11}+h_{22}(1)W_{12}+h_{23}(1)W_{13}\nonumber\\\equiv& L_2^2(W_{21},W_{22},W_{23})+I_2^1(W_{11},W_{12},W_{13})\nonumber\\
Y_{3}(4)=&h_{31}(4)h_{21}^{-1}(4)h_{21}(1)W_{11}+h_{32}(4)h_{22}^{-1}(4)h_{22}(1)W_{12}+h_{33}(4)h_{23}^{-1}(4)h_{23}(1)W_{13}\nonumber\\&+h_{31}(4)h_{11}^{-1}(4)h_{11}(2)W_{21}+h_{32}(4)h_{12}^{-1}(4)h_{12}(2)W_{22}+h_{33}(4)h_{13}^{-1}(4)h_{13}(2)W_{23}.\nonumber
\end{align}
The receiver $R_1$ receives the second linear combination $L_1^2(W_{11},W_{12},W_{13})$ of its desired messages $\{W_{11}, W_{12}, W_{13}\}$ and the interference term $I_1^1(W_{21},W_{22},W_{23})$. However, $R_1$ already receive the same interference at time slot $t=2$ such that $Y_{1}(2)=I_1^1(W_{21},W_{22},W_{23})$. Thus, $L_1^2(W_{11},W_{12},W_{13})$ can be obtained by subtracting $Y_1(2)$ from $Y_1(4)$. Similarly, the receiver $R_2$ can also obtain $L_2^2(W_{21},W_{22},W_{23})$ by subtracting $Y_2(1)$ from $Y_2(4)$. In fact, the received signal $Y_{3}(4)$ is not used in our proposed scheme.\\
At time slot $t=5$, $R_1$ and $R_3$ are P state, i.e., $T_j$ knows $h_{1j}(5)$ and $h_{3j}(5)$. Also, $R_1$ is D state at time slot $t=3$ and $R_3$ is D state at time slot $t=1$, i.e., $T_j$ knows $h_{1j}(3)$ and $h_{3j}(1)$. By using these CSIs, each transmitter transmits the precoded signal at time slot $t=5$ as follows.
\begin{align}
&X_{1}(5)=h_{31}^{-1}(5)h_{31}(1)W_{11}+h_{11}^{-1}(5)h_{11}(3)W_{31}\nonumber\\
&X_{2}(5)=h_{32}^{-1}(5)h_{32}(1)W_{12}+h_{12}^{-1}(5)h_{12}(3)W_{32}\\
&X_{3}(5)=h_{33}^{-1}(5)h_{33}(1)W_{13}+h_{13}^{-1}(5)h_{13}(3)W_{33}.\nonumber
\end{align}
At time slot $t=6$, $R_2$ and $R_3$ are P state, i.e., $T_j$ knows $h_{2j}(6)$ and $h_{3j}(6)$. Also, $R_2$ is D state at time slot $t=3$ and $R_3$ is D state at time slot $t=2$, i.e., $T_j$ knows $h_{2j}(3)$ and $h_{3j}(2)$. By using these CSIs, each transmitter transmits the precoded signal at time slot $t=6$ as follows. 
\begin{align}
&X_{1}(6)=h_{31}^{-1}(6)h_{31}(2)W_{21}+h_{21}^{-1}(6)h_{21}(3)W_{31}\nonumber\\
&X_{2}(6)=h_{32}^{-1}(6)h_{32}(2)W_{22}+h_{22}^{-1}(6)h_{22}(3)W_{32}\\
&X_{3}(6)=h_{33}^{-1}(6)h_{33}(2)W_{23}+h_{23}^{-1}(6)h_{23}(3)W_{33}.\nonumber
\end{align} 
Then, the received signals at time slot $t=5,6$ are also given as
\begin{align}
Y_{1}(5)=&h_{11}(5)h_{31}^{-1}(5)h_{31}(1)W_{11}+h_{12}(5)h_{32}^{-1}(5)h_{32}(1)W_{12}+h_{13}(5)h_{33}^{-1}(5)h_{33}(1)W_{13}\nonumber\\&+h_{11}(3)W_{31}+h_{12}(3)W_{32}+h_{13}(3)W_{33}\nonumber\\\equiv& L_1^3(W_{11},W_{12},W_{13})+I_1^2(W_{31},W_{32},W_{33})\nonumber\\
Y_{2}(5)=&~\text {this~received~signal is not used}\\
Y_{3}(5)=&h_{31}(5)h_{11}^{-1}(5)h_{11}(3)W_{31}+h_{32}(5)h_{12}^{-1}(5)h_{12}(3)W_{32}+h_{33}(5)h_{13}^{-1}(5)h_{13}(3)W_{33}\nonumber\\&+h_{31}(1)W_{11}+h_{32}(1)W_{12}+h_{33}(1)W_{13}\nonumber\\\equiv& L_3^2(W_{31},W_{32},W_{33})+I_3^1(W_{11},W_{12},W_{13})\nonumber\\\nonumber\\
Y_{1}(6)=&~\text {this~received~signal is not used}\nonumber\\
Y_{2}(6)=&h_{21}(6)h_{31}^{-1}(6)h_{31}(2)W_{21}+h_{22}(6)h_{32}^{-1}(6)h_{32}(2)W_{22}+h_{23}(6)h_{33}^{-1}(6)h_{33}(2)W_{23}\nonumber\\&+h_{21}(3)W_{31}+h_{22}(3)W_{32}+h_{23}(3)W_{33}\nonumber\\\equiv& L_2^3(W_{21},W_{22},W_{23})+I_2^2(W_{31},W_{32},W_{33})\\
Y_{3}(6)=&h_{31}(6)h_{21}^{-1}(6)h_{21}(3)W_{31}+h_{32}(6)h_{22}^{-1}(6)h_{22}(3)W_{32}+h_{33}(6)h_{23}^{-1}(6)h_{23}(3)W_{33}\nonumber\\&+h_{31}(2)W_{21}+h_{32}(2)W_{22}+h_{33}(2)W_{23}\nonumber\\\equiv&
L_3^3(W_{31},W_{32},W_{33})+I_3^2(W_{21},W_{22},W_{23}).\nonumber
\end{align}
At time slot $t=5,6$, the receiver $R_1$ obtains the third linear combination $L_1^3(W_{11},W_{12},W_{13})$ by subtracting $Y_1(2)$ from $Y_1(5)$ and the receiver $R_2$ obtains the third linear combination $L_2^3(W_{21},W_{22},W_{23})$ by subtracting $Y_2(3)$ from $Y_2(6)$. Similarly, $R_3$ obtains $L_3^2(W_{31},W_{32},W_{33})$ by subtracting $Y_3(1)$ from $Y_3(5)$ and $L_3^3(W_{31},W_{32},W_{33})$ by subtracting $Y_3(2)$ from $Y_3(6)$. Therefore, in the phase 2, each receiver obtains three linear combinations of the desired messages, i.e., 
\begin{align}
&R_1:\{L_1^1(W_{11},W_{12},W_{13}),L_1^2(W_{11},W_{12},W_{13}),L_1^3(W_{11},W_{12},W_{13})\}\nonumber\\
&R_2:\{L_2^1(W_{21},W_{22},W_{23}),L_2^2(W_{21},W_{22},W_{23}),L_2^3(W_{21},W_{22},W_{23})\}\nonumber\\
&R_3:\{L_3^1(W_{31},W_{32},W_{33}),L_3^2(W_{31},W_{32},W_{33}),L_3^3(W_{31},W_{32},W_{33})\}.\nonumber
\end{align}
As a result, the receiver $R_i$ can decode three desired messages $\{W_{i1},W_{i2},W_{i3}\}$ by solving their three linear equations. Thus, total DoF $9/6=3/2$ is achieved for the 3-user SISO X channel by the proposed achievable scheme.
As you can see in (7), (9), and (10), each transmitter requires the delayed and the perfect CSIs to precode its transmitting messages.
\begin{table}[h]
\caption {Synergistic alternating CSIT states for 3-user X channel}
\centering
\begin{tabular}{|c|c|c|c|c|c|c|} \hline
&\multicolumn{3}{|c|}{Phase 1}&\multicolumn{3}{|c|}{Phase 2}\\\hline
Time&1&2&3&4&5&6\\\hline
\hline$R_1$&N&D&D&P&P&N\\\hline
$R_2$&D&N&D&P&N&P\\\hline
$R_3$&D&D&N&N&P&P\nonumber\\\hline
\end{tabular}
\end{table}\\
\indent Table I summarizes the required synergistic alternating CSIT states at the receiver side. At time slot 1, $R_1$ receives the linear combination of the desired messages $\{W_{11},W_{12},W_{13}\}$, which corresponds to the interference messages to $R_2$ and $R_3$. At time slot 4, $R_1$ receives the second linear combination of the desired messages $\{W_{11},W_{12},W_{13}\}$ and the linear combination of the interference messages by $\{W_{21},W_{22},W_{23}\}$.
At $R_1$, by using the interference signal $I_1^1(W_{21},W_{22},W_{23})$ received at time slot 2, the interference at time slot 4 is cancelled by precoding the transmitted messages as in (7). That is, the precoding is done at each transmitter by using the current channel coefficients (P CSIT) and the previous channel coefficients (D CSIT). Similarly, the interference of $R_1$ at time slot 5 can be cancelled by using the interference at time slot 3 and the precoded transmitted signals at time slot 5 in (9). For $R_2$ and $R_3$, the interference can be cancelled in the similar manner to obtain three linear combinations of the desired messages. As a result, in the phase 1, if the receiver receives the interference, it is used in the phase 2 as D CSIT. Therefore, the receivers should be D state. Also, if the receiver receives the desired signals, the corresponding CSIT is not used in the phase 2. In the phase 2, if the receiver receives linear combinations of its desired messages and interference, it requires P state for precoder and if the receiver receives only the interference, it is in N state as in Table I.\\
\indent It is clear that in the proposed scheme, the CSIT states in Table I can be columnwisely permuted within each phase. Since each phase uses three time slots, there are $3!3!=36$ possible CSIT states with rescheduled transmission order.
\subsection{Achievable Schemes for $M \times N$ SISO X Channel}
In the $M \times N$ SISO X channel in Fig. 2, we have $MN$ independent messages $W_{ij}$, $i=1,\cdot\cdot\cdot,N,~j=1,\cdot\cdot\cdot,M$. Therefore, each receiver requires $M$ linear combinations of $M$ desired messages to recover messages during the given channel uses. Similar to the 3-user case, the proposed achievable schemes are composed of two phases. Depending on the number of users, the following four cases of the achievable schemes are considered as below.
\begin{figure}[h]
\centering
\includegraphics[width=0.6\textwidth]{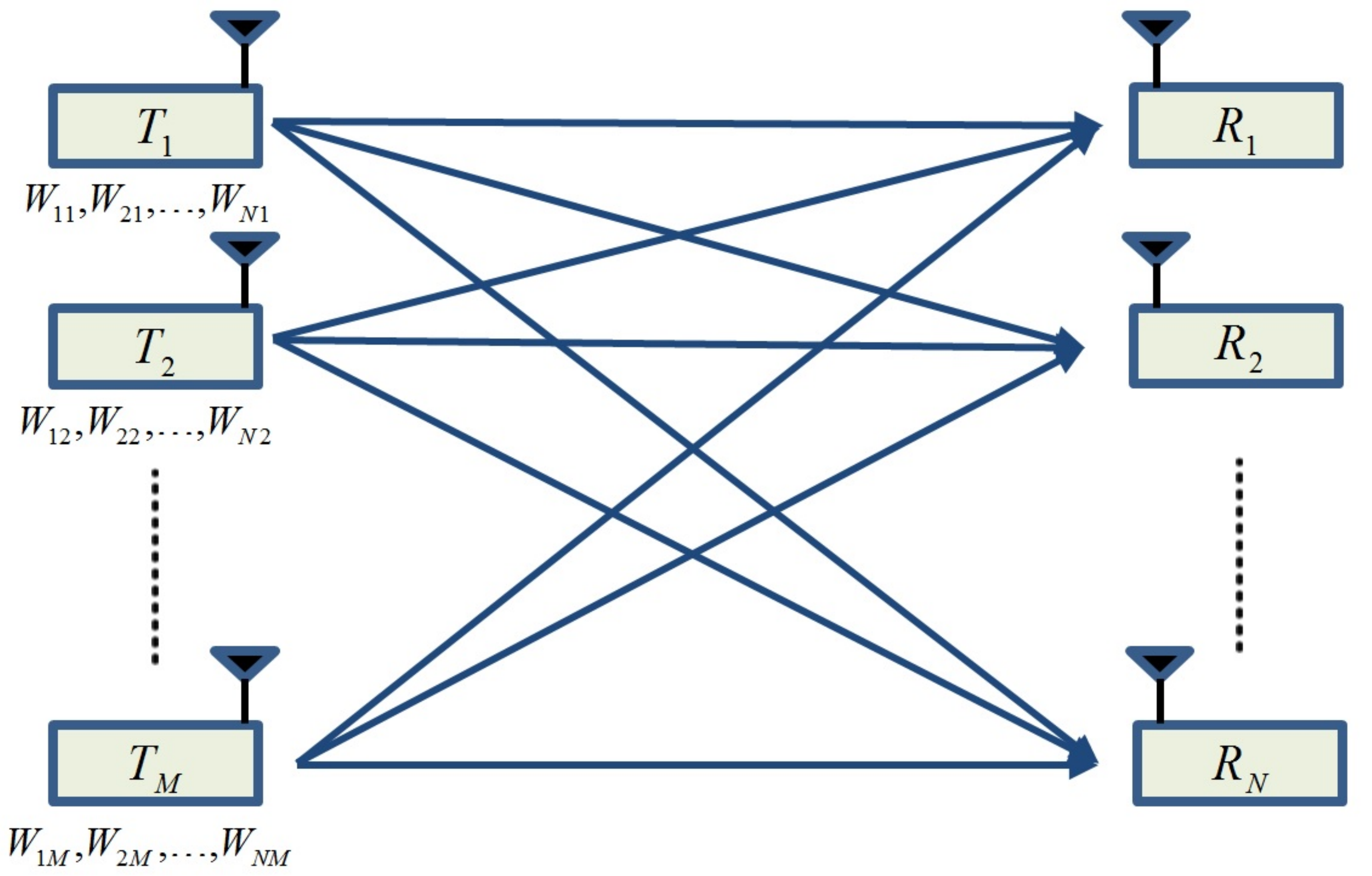}
\caption{$M \times N$ SISO X channel.}
\label{fig:}
\end{figure}

i)~$M\geq N$ except when $M$ is even and $N$ is odd:\\\

\indent\textit{\textbf{Phase 1)}}~~In this phase, $N$ time slots are used. At time slot $i$, all transmitters transmit their desired messages to the receiver $R_i$, i.e., $X_j(i)=W_{ij},~1\leq i\leq N,~1\leq j\leq M$.\\\

\textit{\textbf{Phase 2)}}~~In this phase, all transmitters transmit linear combinations of two desired messages $\{W_{i_1j},W_{i_2j}\}$, $1\leq j\leq M$, to each pair of two receivers $\{R_{i_1},R_{i_2}\}$ at each time slot. In other words, at the time slot $t$, the $j$th transmitter transmits $X_j(t)=f_{i_1j}(t)W_{i_1j}+f_{i_2j}(t)W_{i_2j},~1\leq j\leq M$, where $f_{i_1j}(t)$ and $f_{i_2j}(t)$ are precoding coefficients. Thus, all transmitters transmit ${N \choose 2}$ linear combinations of two desired messages to each pair of two receivers over ${N \choose 2}$ time slots. During these ${N \choose 2}$ time slots, each receiver receives $N-1$ linear combinations of its desired messages and interferences from $M$ transmitters among ${N \choose 2}$ received signals. Since each receiver receives one linear combination of $M$ desired messages from $M$ transmitters in the phase 1, in order to seperate $M$ desired messages, each receiver needs additional $M-1$ linear combinations of $M$ desired messages. For that purpose, all transmitters transmit repeatedly until all receivers have received $M-1$ linear combinations of the desired messages in the phase 2. Let $\alpha$ be the repeated number of ${N \choose 2}$ transmissions in the phase 2. Since each receiver obtains $N-1$ linear combinations of its desired messages during one repetition, the repeated number $\alpha$ should satisfy $\alpha(N-1)=M-1$. 
Then, the total number of transmissions in the phase 2 is equal to $(M-1){N \choose 2}/(N-1)={N(M-1)}/{2}$. The transmitted messages are precoded for interference cancellation using the previous time slot CSI as in (7), (9), and (10).\\
The transmitted signals are given as
\begin{table*}[h]
\centering
\fontsize{10}{16}\selectfont
\begin{tabular}{|c|c|c|c|c|c|c|c|c|c|c|} \hline
&\multicolumn{4}{|c|}{Phase 1}&\multicolumn{6}{|c|}{Phase 2}\\\hline
Time&$1$&$2$&$\cdot\cdot\cdot$&$N$&$N+1$&$N+2$&$\cdot\cdot\cdot$&$N+{N \choose 2}$&$\cdot\cdot\cdot$&$N+\dfrac{N(M-1)}{2}$\\\hline
$\mathbf{Tx}$&$W_1$&$W_2$&$\cdot\cdot\cdot$&$W_N$&$W_1,W_2$&$W_1,W_3$&$\cdot\cdot\cdot$&$W_{N-1},W_N$&$\cdot\cdot\cdot$&$W_{N-1},W_N$\\\hline
\end{tabular} 
\end{table*}\\
where $W_i=\{W_{ij}|j=1,2,...,M\}$ is the desired messages to the receiver $R_i$.\\
Then, the received signals are given as
\begin{table*}[h]
\centering
\fontsize{6}{11}\selectfont
\begin{tabular}{|c|c|c|c|c|c|c|c|c|} \hline
&\multicolumn{4}{|c|}{Phase 1}&\multicolumn{4}{|c|}{Phase 2}\\\hline
Time&$1$&$2$&$\cdot\cdot\cdot$&$N$&$N+1$&$N+2$&$\cdot\cdot\cdot$&$N+\dfrac{N(M-1)}{2}$\\\hline
$R_1$&$L_1^1(W_1)$&$I_1^1(W_2)$&$\cdot\cdot\cdot$&$I_1^{N-1}(W_N)$&$L_1^2(W_1)+I_1^1(W_2)$&$L_1^3(W_1)+I_1^2(W_3)$&$\cdot\cdot\cdot$&$-$\\\hline
$R_2$&$I_2^1(W_1)$&$L_2^1(W_2)$&$\cdot\cdot\cdot$&$I_2^{N-1}(W_N)$&$L_2^2(W_2)+I_2^1(W_1)$&$-$&$\cdot\cdot\cdot$&$-$\\\hline
$R_3$&$I_3^1(W_1)$&$I_3^2(W_2)$&$\cdot\cdot\cdot$&$I_3^{N-1}(W_N)$&$-$&$L_3^2(W_3)+I_3^1(W_1)$&$\cdot\cdot\cdot$&$-$\\\hline
$\cdot\cdot\cdot$&$\cdot\cdot\cdot$&$\cdot\cdot\cdot$&$\cdot\cdot\cdot$&$\cdot\cdot\cdot$&$\cdot\cdot\cdot$&$\cdot\cdot\cdot$&$\cdot\cdot\cdot$&$\cdot\cdot\cdot$\\\hline
$R_{N-1}$&$I_{N-1}^1(W_1)$&$I_{N-1}^2(W_2)$&$\cdot\cdot\cdot$&$I_{N-1}^{N-1}(W_N)$&$-$&$-$&$\cdot\cdot\cdot$&$L_{N-1}^M(W_{N-1})+I_{N-1}^{N-1}(W_N)$\\\hline
$R_N$&$I_N^1(W_1)$&$I_N^2(W_2)$&$\cdot\cdot\cdot$&$L_N^{1}(W_N)$&$-$&$-$&$\cdot\cdot\cdot$&$L_{N}^M(W_{N})+I_{N}^{N-1}(W_{N-1})$\\\hline
\end{tabular} 
\end{table*}\\
where `-' means useless received interference signal.\\
\indent In the phase 1, each receiver obtains one linear combination of the desired messages. In the phase 2, each receiver obtains $M-1$ additional linear combinations of its desired messages by subtracting the received interference at the phase 1. For example, $R_1$ obtains $L_1^2(W_1)$ by substracting $I_1^1(W_2)$ at time slot 2 from $L_1^2(W_1)+I_1^1(W_2)$ at time slot $N+1$. Using $N+N(M-1)/2$ channel uses, each receiver has $M$ linear combinations of its desired messages, i.e.,
\begin{align}
&R_1:\{L_1^1(W_{1}),L_1^2(W_{1}),\cdot\cdot\cdot,L_1^M(W_{1})\}\nonumber\\
&R_2:\{L_2^1(W_{2}),L_2^2(W_{2}),\cdot\cdot\cdot,L_2^M(W_{2})\}\nonumber\\
&~~~~~~~~~~~~~~~~~~~~~~\cdot\cdot\cdot\nonumber\\
&R_N:\{L_N^1(W_{N}),L_N^2(W_{N}),\cdot\cdot\cdot,L_N^M(W_{N})\}.\nonumber
\end{align} 
Therefore, each receiver can decode $M$ desired messages by solving $M$ linear equations and the total DoF ${MN}/(N+N(M-1)/2)={2M}/(M+1)$ is achieved by the proposed scheme.\\
\begin{table*}[h]
\centering
\caption{Synergistic alternating CSIT states for $M\times N$ SISO X channel}
\begin{tabular}{|c|c|c|c|c|c|c|c|c|c|} \hline
&\multicolumn{5}{|c|}{Phase 1}&\multicolumn{4}{|c|}{Phase 2}\\\hline
Time&$1$&$2$&3&$\cdot\cdot\cdot$&$N$&$N+1$&$N+2$&$\cdot\cdot\cdot$&$N+\dfrac{N(M-1)}{2}$\\\hline
$R_1$&N&D&D&$\cdot\cdot\cdot$&D&P&P&$\cdot\cdot\cdot$&N\\\hline
$R_2$&D&N&D&$\cdot\cdot\cdot$&D&P&N&$\cdot\cdot\cdot$&N\\\hline
$R_3$&D&D&N&$\cdot\cdot\cdot$&D&N&P&$\cdot\cdot\cdot$&N\\\hline
$\cdot\cdot\cdot$&$\cdot\cdot\cdot$&$\cdot\cdot\cdot$&$\cdot\cdot\cdot$&$\cdot\cdot\cdot$&$\cdot\cdot\cdot$&$\cdot\cdot\cdot$&$\cdot\cdot\cdot$&$\cdot\cdot\cdot$&$\cdot\cdot\cdot$\\\hline
$R_{N-1}$&D&D&D&$\cdot\cdot\cdot$&D&N&N&$\cdot\cdot\cdot$&P\\\hline
$R_N$&D&D&D&$\cdot\cdot\cdot$&N&N&N&$\cdot\cdot\cdot$&P\\\hline
\end{tabular}
\end{table*}
Synergistic alternating CSIT states used in the proposed scheme are listed in Table II. Similar to the 3-user case, in the phase 1, the receiver that receives the interference has D state to use CSIT in the phase 2 and the receiver that receives the desired signals has N state. In the phase 2, the receiver that receives linear combinations of its desired messages and interferences has P state for precoder and the receiver that receives only the interferences has N state as in Table II.\\ 
\indent In the proposed scheme, the CSIT states in Table II can be columnwisely permuted among columns within each phase and there are $N!(N(M-1)/2)!$ possible CSIT states with rescheduled transmission order.\\\

ii)~$M\geq N$, when $M$ is even and $N$ is odd:\\\
\indent In this case, the phase 1 is the same as that in the previous case but there are some modifications in the phase 2. For even $M$ and odd $N$ case, $(M-1){N \choose 2}/(N-1)=N(M-1)/2$ cannot be an integer. To achieve the same DoF as the previous case, the number of desired messages and the channel uses are doubled, i.e., $W_i^k=\{W_{ij}^k|j=1,2,...,M,~k=1,2\}$ is the desired messages for the receiver $R_i$ and $2(N+N(M-1)/2)$ is the number of channel uses.\\ 
The transmitted signals are given as
\begin{table*}[h]
\centering
\fontsize{10}{16}\selectfont
\begin{tabular}{|c|c|c|c|c|c|c|c|c|c|c|} \hline
&\multicolumn{4}{|c|}{Phase 1}&\multicolumn{6}{|c|}{Phase 2}\\\hline
Time&$1$&$2$&$\cdot\cdot\cdot$&$2N$&$2N+1$&$2N+2$&$\cdot\cdot\cdot$&$2(N+{N \choose 2})$&$\cdot\cdot\cdot$&$2(N+\dfrac{N(M-1)}{2})$\\\hline
$\mathbf{Tx}$&$W_1^1$&$W_2^1$&$\cdot\cdot\cdot$&$W_N^2$&$W_1^1,W_2^1$&$W_1^1,W_3^1$&$\cdot\cdot\cdot$&$W_{N-1}^2,W_N^2$&$\cdot\cdot\cdot$&$W_{N-1}^2,W_N^2$\\\hline
\end{tabular} 
\end{table*}\\
where $W_i^k=\{W_{ij}^k|j=1,2,...,M,~k=1,2\}$ is the desired message to the receiver $R_i$.\\
\indent All other procedures are the same as the previous case. For convenience, the precoding strategy and CSIT states are skipped. After the phase 2, during $2(N+N(M-1)/2)$ channel uses, each receiver has $2M$ linear combinations of its desired messages. Therefore, the $N$ receivers can decode $2M$ desired messages by solving $2M$ linear equations and total DoF $2MN/2(N+N(M-1)/2)=2M/(M+1)$ is achieved by the proposed scheme.\\\

iii)~$N\geq M$ and even $N$:\\\
\indent The overall situation of the scheme is the same but we consider minor modification in the phase 2. Achievable scheme is also composed of two separate phases described as below.\\

\indent\textit{\textbf{Phase 1)}}~~In this phase, $N$ time slots are used. At time slot $i$, all transmitters transmit their desired messages to the receiver $R_i$, i.e., $X_j(i)=W_{ij},~1\leq i\leq N,~1\leq j\leq M$.\\\

\textit{\textbf{Phase 2)}}~~In this phase, the transmitters send desired message pairs to the corresponding receiver pairs, where all pairs of desired messages assume to be disjoint. Thus, during $N/2$ time slots, each receiver receives one linear combination of its desired messages and interferences. After the phase 1, each receiver obtains one linear combination of the desired messages. Therefore, each receiver needs $M-1$ additional linear combinations of $M$ desired messages. For that purpose, all transmitters transmit repeatedly until all receivers have received $M-1$ linear combinations of the desired messages in the phase 2. Let $\alpha$ be the repeated number of $N/2$ transmissions in the phase 2. Since each receiver obtains one linear combinations of its desired messages during one repetition, the repeated number $\alpha$ should be $\alpha=M-1$. 
Then, the total number of transmissions in the phase 2 is equal to $(M-1)N/2$. The transmitted messages are precoded for interference cancellation using the previous time slot CSI.\\ 
The transmitted signals are given as
\begin{table*}[h]
\centering
\fontsize{10}{16}\selectfont
\begin{tabular}{|c|c|c|c|c|c|c|c|c|c|c|} \hline
&\multicolumn{4}{|c|}{Phase 1}&\multicolumn{6}{|c|}{Phase 2}\\\hline
Time&$1$&$2$&$\cdot\cdot\cdot$&$N$&$N+1$&$N+2$&$\cdot\cdot\cdot$&$N+\dfrac{N}{2}$&$\cdot\cdot\cdot$&$N+(M-1)\dfrac{N}{2}$\\\hline
$\mathbf{Tx}$&$W_1$&$W_2$&$\cdot\cdot\cdot$&$W_N$&$W_1,W_2$&$W_3,W_4$&$\cdot\cdot\cdot$&$W_{N-1},W_N$&$\cdot\cdot\cdot$&$W_{N-1},W_N$\\\hline
\end{tabular} 
\end{table*}\\
where $W_i=\{W_{ij}|j=1,2,...,M\}$ is the desired messages for the receiver $R_i$.\\
\indent All other procedures are the same as the previous case. For convenience, the precoding strategy and CSIT states are skipped. After the phase 2, during $N+(M-1)N/2$ channel uses, each receiver has $M$ linear combinations of its desired messages. Therefore, the $N$ receivers can decode $M$ desired messages by solving $M$ linear equations and DoF $MN/(N+(M-1)N/2)=2M/(M+1)$ is achieved by the proposed scheme.\\\

iv)~$N\geq M$ and odd $N$:\\\
\indent Similar to the case ii), the number of desired messages and channel uses are doubled, i.e., $W_i^k=\{W_{ij}^k|j=1,2,...,M,~k=1,2\}$ is the desired messages for the receiver $R_i$ and $2(N+(M-1)N/2)$ is the number of channel uses.\\
The transmitted signals are given as
\begin{table*}[h]
\centering
\fontsize{10}{16}\selectfont
\begin{tabular}{|c|c|c|c|c|c|c|c|c|c|c|} \hline
&\multicolumn{4}{|c|}{Phase 1}&\multicolumn{6}{|c|}{Phase 2}\\\hline
Time&$1$&$2$&$\cdot\cdot\cdot$&$2N$&$2N+1$&$2N+2$&$\cdot\cdot\cdot$&$2(N+\dfrac{N}{2})$&$\cdot\cdot\cdot$&$2(N+(M-1)\dfrac{N}{2})$\\\hline
$\mathbf{Tx}$&$W_1^1$&$W_2^1$&$\cdot\cdot\cdot$&$W_N^2$&$W_1^1,W_2^1$&$W_3^1,W_4^1$&$\cdot\cdot\cdot$&$W_{N-1}^2,W_N^2$&$\cdot\cdot\cdot$&$W_{N-1}^2,W_N^2$\\\hline
\end{tabular} 
\end{table*}\\
where $W_i^k=\{W_{ij}^k|j=1,2,...,M,~k=1,2\}$ is the desired messages to the receiver $R_i$.\\  
\indent All other procedures are the same as the previous case. For convenience, the precoding strategy and CSIT states are skipped. After the phase 2, during $2(N+(M-1)N/2)$ channel uses, each receiver has $2M$ linear combinations of its desired messages. Therefore, the $N$ receivers can decode $2M$ desired messages by solving $2M$ linear equations and DoF $2MN/2(N+(M-1)N/2)=2M/(M+1)$ is achieved by the proposed scheme.\\
\indent As a result, the proposed schemes achieve $2M/(M+1)$ DoF for $M \times N$ SISO X channel with synergistic alternating CSIT.

\section{Conclusion}
We propose the achievable scheme of DoF $2M/(M+1)$ for $M\times N$ SISO X channel with synergistic alternating CSIT. The synergistic benefits for the alternating CSIT still exist in the $M\times N$ SISO X channel. Total DoF $2M/(M+1)$ implies that if the number of transmitters is fixed to $M$, the total DoF is determined by $M$ and each receiver has DoF divided by the total number of receivers. In the proposed scheme for the $M\times N$ SISO X channel, the achievable DoF for $M\times N$ X channel does not scale with the number of users unlike the DoF with the perfect and instantaneous CSIT.

\section{Acknowledgment}
This paper was supported by the ICT Collaboration Program of Samsung Electronics.

\end{document}